\documentclass[aps,prl,reprint,superscriptaddress,longbibliography]{revtex4-2}
\usepackage{physics}
\usepackage{graphicx}
\usepackage{bm} 
\usepackage[table]{xcolor}
\usepackage{hyperref}
\usepackage{bigints}
\usepackage{booktabs}

\definecolor{myviolet}{RGB}{232,151,232}
\definecolor{lightpink}{RGB}{232,211,212}

\usepackage{accents}

\newcommand{\reffig}[1]{Fig.~\ref{#1}}
\newcommand{\reffigs}[2]{Figs.~\ref{#1}-\ref{#2}}
\newcommand{\refeq}[1]{Eq.~(\ref{#1})}
\newcommand{\refeqs}[2]{Eqs.~(\ref{#1})-(\ref{#2})}

\newcommand{\lT}{\mathbin{\mathrm{T}}}

\usepackage{tcolorbox}

\usepackage{varwidth}

\begin{document}

\title{A Theory of Nested Cascading in Directed Logic}


\author{Ihar Babushkin}

\affiliation{ Institute of  Quantum Optics, Leibniz University Hannover,
  Welfengarten 1, 30167 Hannover, Germany}

\affiliation{Max Born Insitute, Berlin, Germany}

\affiliation{Cluster of Excellence PhoenixD (Photonics, Optics, and
  Engineering - Innovation Across Disciplines), Leibniz University Hannover, Welfengarten 1, 30167  Hannover, Germany}

\email{babushkin@iqo.uni-hannover.de}

\author{Oliver Melchert}

\affiliation{ Institute of  Quantum Optics, Leibniz University Hannover,
  Welfengarten 1, 30167 Hannover, Germany}

\affiliation{Cluster of Excellence PhoenixD (Photonics, Optics, and
  Engineering - Innovation Across Disciplines), Leibniz University Hannover, Welfengarten 1, 30167  Hannover, Germany}

\author{Ayhan Demircan}

\affiliation{ Institute of  Quantum Optics, Leibniz University Hannover,
  Welfengarten 1, 30167 Hannover, Germany}

\affiliation{Cluster of Excellence PhoenixD (Photonics, Optics, and
  Engineering - Innovation Across Disciplines), Leibniz University Hannover, Welfengarten 1, 30167  Hannover, Germany}

\author{Uwe Morgner}

\affiliation{ Institute of  Quantum Optics, Leibniz University Hannover,
  Welfengarten 1, 30167 Hannover, Germany}

\affiliation{Cluster of Excellence PhoenixD (Photonics, Optics, and
  Engineering - Innovation Across Disciplines), Leibniz University Hannover, Welfengarten 1, 30167  Hannover, Germany}

\begin{abstract}
  In directed logic (DL), electronically controlled optical elements
  serve as photonic gates. Such electro-optical elements are of mixed
  nature: they have two inputs—electronic and optical—but only one,
  optical, output. This makes cascading such gates without repeated
  conversion between optical and electronic representations
  cumbersome. This problem can be largely overcome using the nested
  cascading scheme proposed by Shamir and Hardy [Opt. Express, 17, 150
  (2007)]. Although promising as a solution to the cascading problem,
  the Shamir–Hardy scheme has so far been proved or implemented only
  for a few simplest cases. Here, we develop a general rigorous theory
  of nested cascading, valid for an arbitrary number of gates. We
  propose a variant of the algorithm which easily extendable to large
  number of gates, and rigorously prove its validity. Furthermore,
  we analyze how nested cascading scales with the size of the
  corresponding Boolean formula. We show that good (linear)
  scalability is guaranteed in many important cases, while the average
  scaling with respect to the corresponding Boolean formulas is only
  moderately polynomial, with an exponent of approximately 3/2. Yet,
  the worst-case scaling remains exponential with respect to more
  general Boolean circuits allowing sharing and reuse of intermediate
  results.
\end{abstract}




\maketitle

\section{Introduction}

Electronic logical switches and computing devices in general are known
to consume large amounts of energy per operation compared to
other computational architectures--for instance, the human brain. Besides, today's
commonly used architectures suffer from the ``von Neumann bottleneck''
-- the inherent conflict between data processing and data
transfer to and from memory, resulting from the fundamentally
sequential nature of such architectures. Popular approaches
to overcoming these problems include delegating particular sub-tasks to
specialized units, such as graphics processing units (GPUs),
hard-coded neural networks, or specialized arithmetic logic units
(ALUs).

\begin{figure*}[thpb] 
  \center{\includegraphics[width=0.7\textwidth]{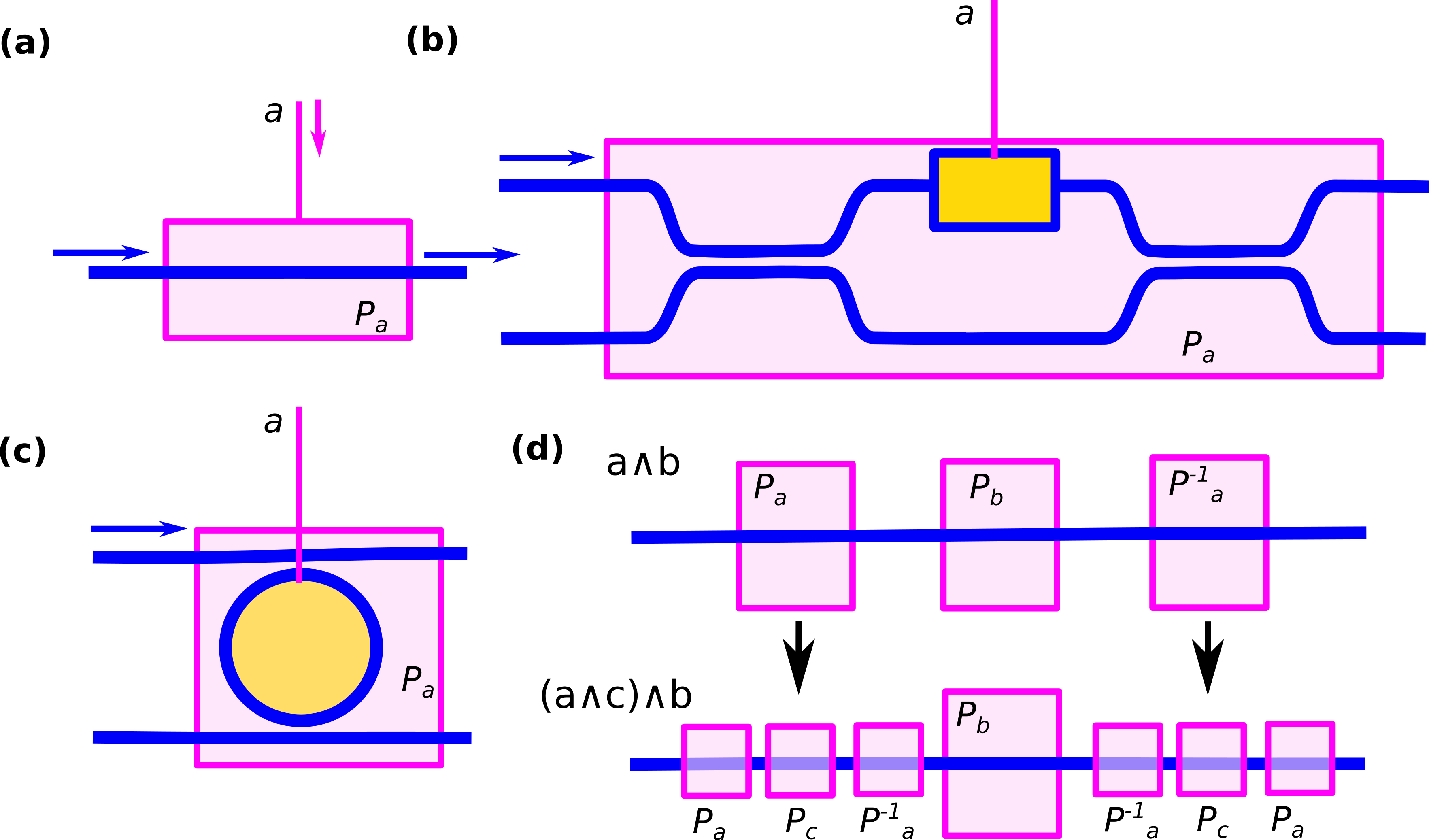}}
  \caption{(a) A basic element of DL, containing two inputs [optical
    (blue incoming arrow) and electronic (pink arrow)] and one
    optical output (blue outgoing arrow). The logical value $a$,
    supplied via the electronic channel, controls the optical path,
    thereby realizing an $a$-dependent linear operator $P_a$. (b,c)
    Exemplary realizations of a DL element $P_a$ using a Mach--Zehnder
    interferometer (b) or a cavity (c). Yellow rectangles denote
    electro-optical or thermo-optical elements, controlled by the
    electronic input $a$. (d) General idea of nested cascading
    suggested by Shamir and Hardy in \cite{hardy07}, shown by example.
    An exemplary logical gate $a\land b$ contains three basic DL
    elements: $P_a$, $P_b$, and $P_a^{-1}$ (the inverse of $P_a$), and
    an exemplary cascading step
    $a\land b\to (a\land c)\land b$. In the cascading procedure, we
    replace $P_a$ with the nested definition of $a\land c$ (see
    \reffig{fig:cascading} for details).
    }
\label{fig:DL}
\end{figure*}

On the other hand, during the last few years, large-scale programmable photonic
integrated circuits (PICs)
\cite{bogaerts20,ying20,tang21,zhou22,ning24,perez25,bente25} have been developed. Such devices
can often even be integrated directly into electronic circuits using
established CMOS technology \cite{sun15,atabaki18,bogaerts20,bente25}.
At the same time, rapid progress in time-varying photonics
[TVM]—including temporal cavity engineering \cite{babushkin26},
programmable spatio-temporal media \cite{demircan26}, and all-optical
control \cite{demircan11,husko11,langford11,jandieri20,yanagimoto22,babushkin22,li24,yanagimoto26,dai26} has considerably expanded the range of dynamic optical functionalities available for photonic information processing.

One of these
approaches is based (see \reffig{fig:DL}) on optical elements such as couplers or cavities,
forming a multi-port optical network. Every one of these optical elements
is electronically controlled using, for instance,
electro-optic or thermo-optic effects. In this way, the
electronic signal reconfigures an otherwise linear optical network,
allowing it to perform logical operations
\cite{caulfield10,shamir86,hardy07,caulfield11,tian11a,ding17,xiao18,feng21,tang21,xiao21,sharma21,qiu22,yuan23,alkhazragi24,wang24}
or directly perform matrix multiplications \cite{xu22,zhou22}. This approach
has already proved its ability to overcome the overall power consumption and the
dissipated power density limits compared to standard electronic
units for clock frequencies beyond $\sim$10~GHz \cite{ying20}.

The circuits performing logical operations in such a way are often
gathered under the term ``directed logic'' (DL) \cite{xiao21,qiu22,yuan23,wang24,alkhazragi24}.
On the level of a single or a few elements,
multiple physical realizations of DL have been suggested,
using, as a basic element, a Mach-Zehnder interferometer  \reffig{fig:DL}b
\cite{ding17,xiao21} or cavities \reffig{fig:DL}c coupled to propagation lines
\cite{tian11a,xiao18,yuan23,alkhazragi24,wang24}, electronically controlled using electro-optic
\cite{ding17} or thermo-optic \cite{ding17,xiao18,xiao21,yuan23,alkhazragi24,wang24} effects.

When trying to connect such gates, we must remember that DL gates
must be considered as devices with not only an optical input and output,
but, in addition, also an electronic input (see \reffig{fig:DL}a), the electronic signal
determining the path taken by light, as mentioned above. The big
advantage of DL is that, as soon as the electronic input is set, the
processing of the optical signal takes place ``with the speed of
light'' and, potentially, as long as losses are small, with
nearly zero energy consumption \cite{hardy07,caulfield10}. However,
because the input contains both electronic and optical parts, whereas
the output is only optical, cascading DL gates (that is, directing
the output of one gate to the input of another) cannot be
straightforward: for this, one needs to connect devices of different
physical nature, electronic and optical. Of course, this can be
achieved by converting the optical output into an electronic
signal, but this operation is relatively slow and energy-intensive,
and thus diminishes most of the potential power of DL. There are many
possible variants of partial cascadability that work in
particular cases; some of them are discussed in \cite{caulfield07}.
For instance, DL elements allow one to scalably calculate conjunctive normal
forms (CNFs). Another example of partial cascading is the realization
of an adder \cite{ying20}, where bitwise addition was performed
electronically in parallel, whereas the carry bit was calculated
using DL.

These examples put forward the following question: suppose we have an
arbitrary Boolean function of $a,b,c,\ldots$ with $n$ variables,
realized using logical gates such as OR or AND. 
Can we combine the corresponding optical DL gates into
an optical circuit, so that, after we set all the electronic inputs
$a,b,c,\ldots$ (and their negations
$\overline{a},\overline{b},\overline{c},\ldots$) of our DL elements,
the circuit produces the required Boolean function at the output? This
is probably the strongest form of cascadability one can request from
DL. In \cite{hardy07}, Shamir and Hardy suggested a nested cascading
approach to address this problem. In this approach, at every nesting
step, a DL gate is replaced by a whole DL circuit implementing this
gate (see \reffig{fig:DL}d), giving rise to a recursive procedure.
Yet, Shamir and Hardy's design for the nested-cascadable elements was
rather cumbersome, denying a straightforward extension to an arbitrary
number of gates; besides, it has been tested only on a few-gate
configurations, and a rigorous justification and proof for arbitrarily
complex cases is missing up to now.

In this article, we close this gap by developing a general theory of
nested cascading. We prove rigorously that the main claim of Shamir
and Hardy is indeed valid: one can construct a nested circuit
implementing an arbitrary Boolean formula. We note that, whereas AND
and OR do not form a universal gate set, in the DL paradigm it is as
easy to set the logical variable $a$ as its negation
$\overline{a}$. Once the negated variables are available, AND and OR
gates alone are sufficient to represent any Boolean formula.
Consequently, although the original proposal \cite{hardy07} included a
larger set of gates, in this article we focus exclusively on AND and
OR gates.

Furthermore, we consider the question of the scalability of nested
cascading with the size of the corresponding Boolean expression. In
nested logic, one of the two input variables is recursively
duplicated, leading to a potentially exponential blowup. Whereas the
naive approach to this question indeed suggests such a behavior
\cite{caulfield07,miller10a}, the question is quite nontrivial because
of the complex nature of nested cascading. We consider this question
in detail, showing that both the worst-case and the typical scaling
follow not an exponential but a power-law dependence with an exponent
of $3/2$ with respect to the corresponding Boolean formula. Moreover,
for a large class of Boolean expressions, the number of DL gates grows
only linearly with the number of logical gates.

\begin{figure*}[thpb] 
  \center{\includegraphics[width=0.7\textwidth]{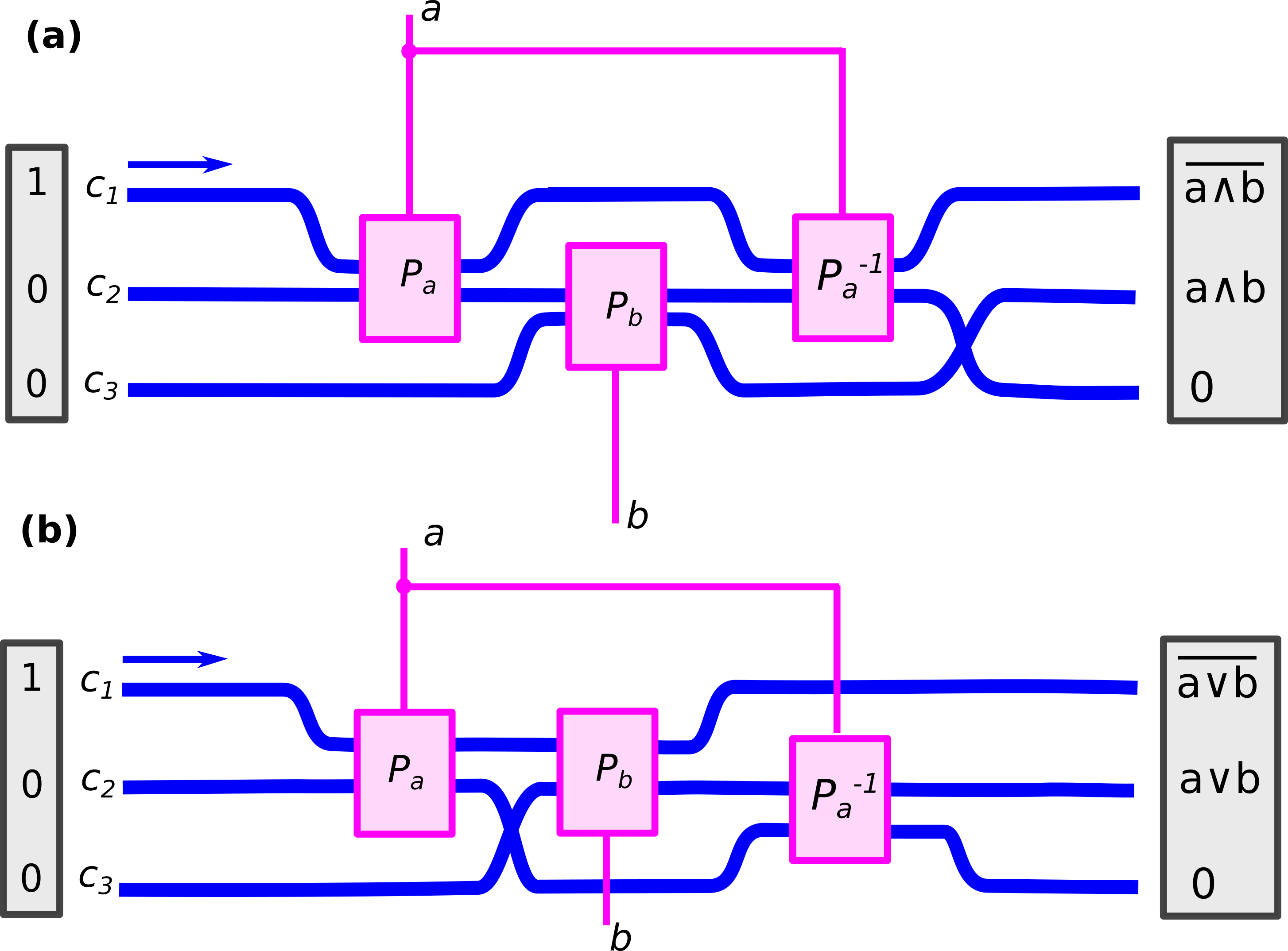}} 
  \caption{Basic DL gates, representing AND (a) and OR (b). Blue lines
    show the optical paths, pink lines indicate the paths for the
    electronic signals, and pink regions show electro-optical
    coupling. Numbers in the gray-shaded box indicate the corresponding input and
output channels when these gates are used as standalone gates (rather
than as part of a larger DL circuit). Up to the channel numbering,
these gates are equivalent to those in \cite{hardy07}. }
\label{fig:base}
\end{figure*}

\section{Nested cascading}
\label{sec:directed-logic}

\subsection{Formulating the procedure}
\label{sec:form-proc}

In the nested cascading, suggested by Shamir and Hardy \cite{hardy07},
the basic building blocks of the gates AND ($a\land b$) and OR
($a\lor b$) are the linear operators $P_{a}$, $P_{b}$ and their
inverses $P^{-1}_{a}$, $P^{-1}_{b}$, acting on a pair of channels (see
\reffig{fig:base}(a,b), where they are represented by pink boxes),
whose action is controlled by the logical variables $a$, $b$, so that
for $a=0$ the operator does nothing,
whereas for $a=1$ it swaps the two signals:
\begin{gather}
  \label{eq:P0}
  P_{0}=\mathrm{Identity}, \\
  P_{1}=\mathrm{Swap},
  \label{eq:P1}
\end{gather}
Obviously, $P^{-1}_a=P_a$.
Technically, $P$ can be realized by couplers or cavities, controlled using an
electronic signal representing $a$ or $b$ via electro-optic
\cite{ding17} or thermo-optic
\cite{ding17,xiao18,xiao21,yuan23,alkhazragi24,wang24} effects [see \reffig{fig:DL}(b,c)].

The gates $\land$ and $\lor$ presented in \reffig{fig:base}(a,b) are
equivalent to those suggested by Shamir and Hardy in \cite{hardy07},
even if they look somewhat different (see Supplementary for a more
direct comparison). Every gate, after the electronic input is set,
represents a linear optical network with 3 inputs and 3 outputs
($3\times3$ optical multiport). Every
gate assumes the optical input to be $(1,0,0)$ [the optical signal
strength denoted by $1$ can be arbitrary, but must, of course, be the
same for all gates]. The output vector has the same energy as the
input, which is a consequence of unitarity.

Depending on the values of $a$ and $b$, the optical signal from
channel~1 is redirected either to the first output channel [this
happens if $\overline{a\land b}=1$ for the AND gate in
\reffig{fig:base}(a) and if $\overline{a\lor b}=1$ for the OR gate in
\reffig{fig:base}(b)] or to one of the two remaining channels in the
opposite case. That means that, if we look at the first output, it
``calculates'' the logical expression $\overline{a\land b}$ or
$\overline{a\lor b}$. Other outputs calculate different logical
expressions---the full truth tables for all outputs are given in
Appendix~A (\reffig{fig:and_or}).

\begin{figure*}[htpb] 
  \center{\includegraphics[width=0.7\textwidth]{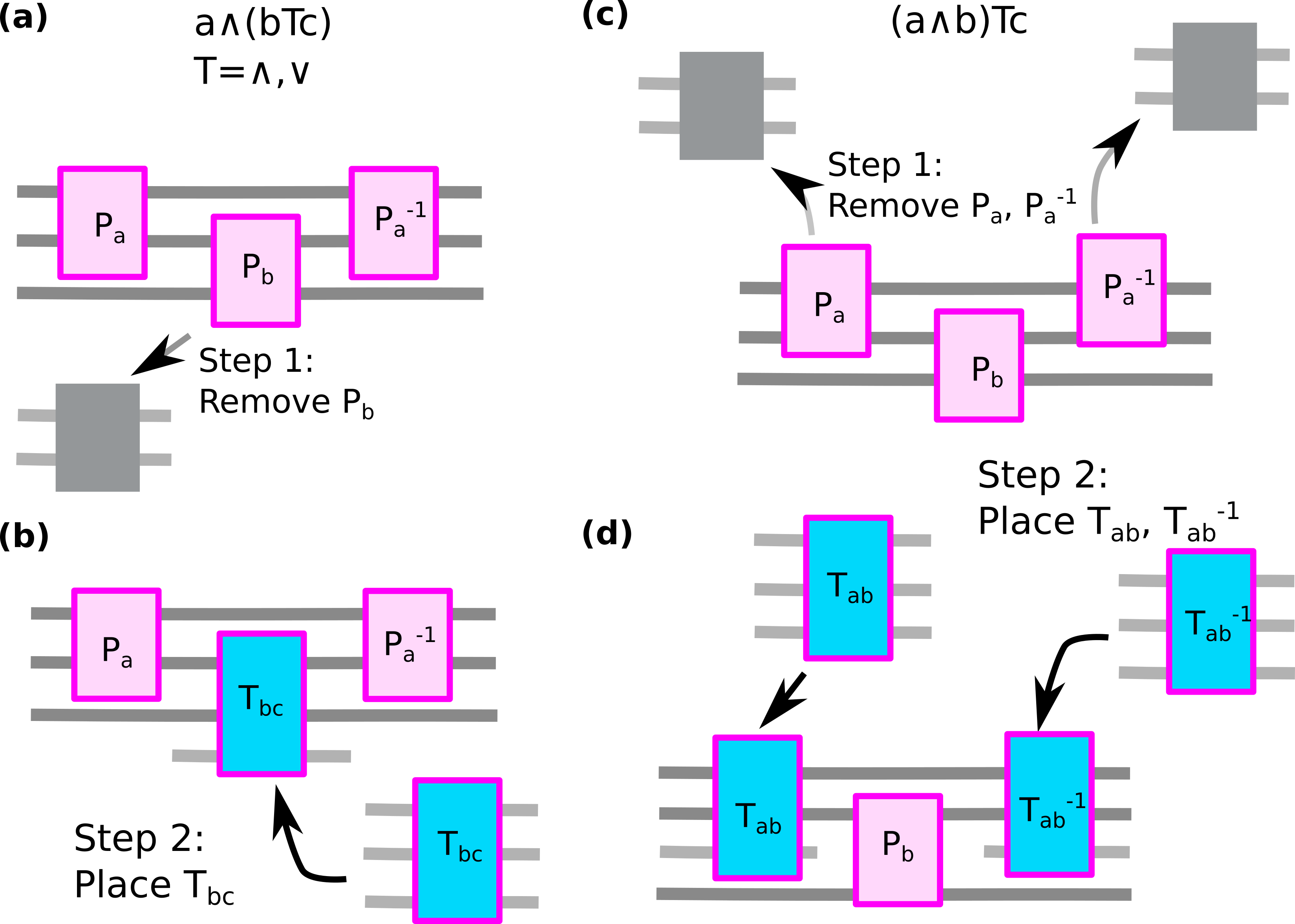}} 
  \caption{Nested cascading of DL gates \cite{hardy07}. The cascading
  procedure applied at the lowest level of the circuit to
  obtain $a\land b\to a\land (b\lT c)$ (a,b), where
  $\lT\in\{\land,\lor\}$, or $a\land b\to (a \lT b)\land c$ (c,d). The last input and output channels are always left unused, while the
remaining two inputs and outputs are connected in an order-preserving
manner.
  An analogous procedure is valid for $a\lor b\to a\lor (b\lT c)$ and
  $a\lor b\to (a\lT b)\lor c$. For simplicity, we denote
  $P_{x\lT y}$ by $T_{xy}$ for $x,y\in\{a,b,c\}$. Note that the last
  slot in every gate must be replaced by the inverse operator
  $P^{-1}_{x\lT y}=T^{-1}_{xy}$.}
\label{fig:cascading}
\end{figure*}

In \cite{hardy07}, furthermore, it was described how these basic gates
can be combined with each other, representing arbitrary Boolean
formulas. To do this, Hardy and Shamir suggested a nested approach. To
create the gate $(a \lor c) \land b$, one should take the DL circuit
for $a \land b$ and replace $P_a$ with the
DL circuit for $a \lor c$ and $P_a^{-1}$ with the DL circuit for
$a \lor c$, only reversed (see \reffig{fig:cascading}a,b). To obtain
$c \land (a \lor b)$, they suggested taking the DL circuit for $c \land b$
and replacing $P_b$ with $a \lor b$ (see \reffig{fig:cascading}c,d).

Despite a very promising proposal, the authors checked their approach
only on a few examples. They also did not provide any proof or truth
tables. Besides, even if we assume it works, from the way they
presented the circuits it is difficult to understand how exactly they
can be scaled to a larger number of gates. Indeed, the procedure
suggests that we replace a 2-port device $P_a$, $P_b$ with 3-port
devices (representing AND or OR). This replacement leaves one of the
inputs and one of the outputs ``hanging'' (see
\reffig{fig:cascading}), and from the discussion in \cite{hardy07} it
is not clear at all which of the inputs and which of the outputs it
should be. Moreover, the orverall order of the inputs and outputs by
this substitution is not fixed in \cite{hardy07}. Here, we adopt a
gate design for which the corresponding rules are particularly simple:
the last input and output channels are always left unused, while the
remaining two inputs and outputs are connected in an order-preserving
manner, with the lower-numbered input connected to the lower-numbered
output. We note that the resulting circuits are not always
topologically equivalent to those presented in Ref.~\cite{hardy07}
(see Supplementary Information).
\begin{figure*}[htpb] 
  \center{\includegraphics[width=0.5\textwidth]{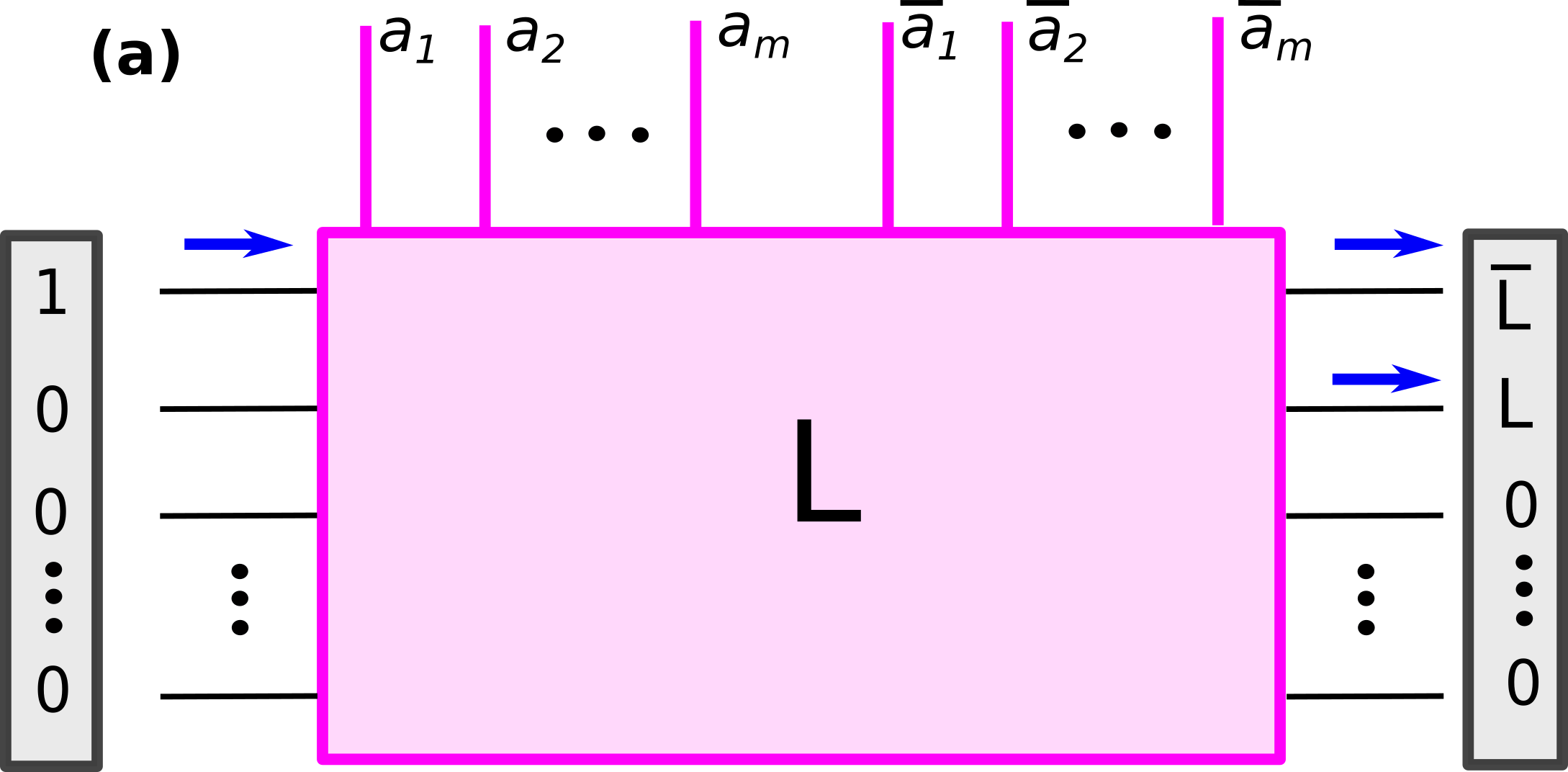}} 
  \caption{A general DL circuit representing
$L(a_1,\overline{a}_1,a_2,\overline{a}_2,\ldots)$ after cascading.
After the variables $a_i$ and $\overline{a}_i$ are set, the entire
circuit becomes a linear operator that transforms an input vector into
an output vector. The gray boxes show the valid input and the
corresponding output vectors. }
\label{fig:cascading-gen}
\end{figure*}

\begin{figure*}[ht!]
  \center{\includegraphics[width=1\linewidth]{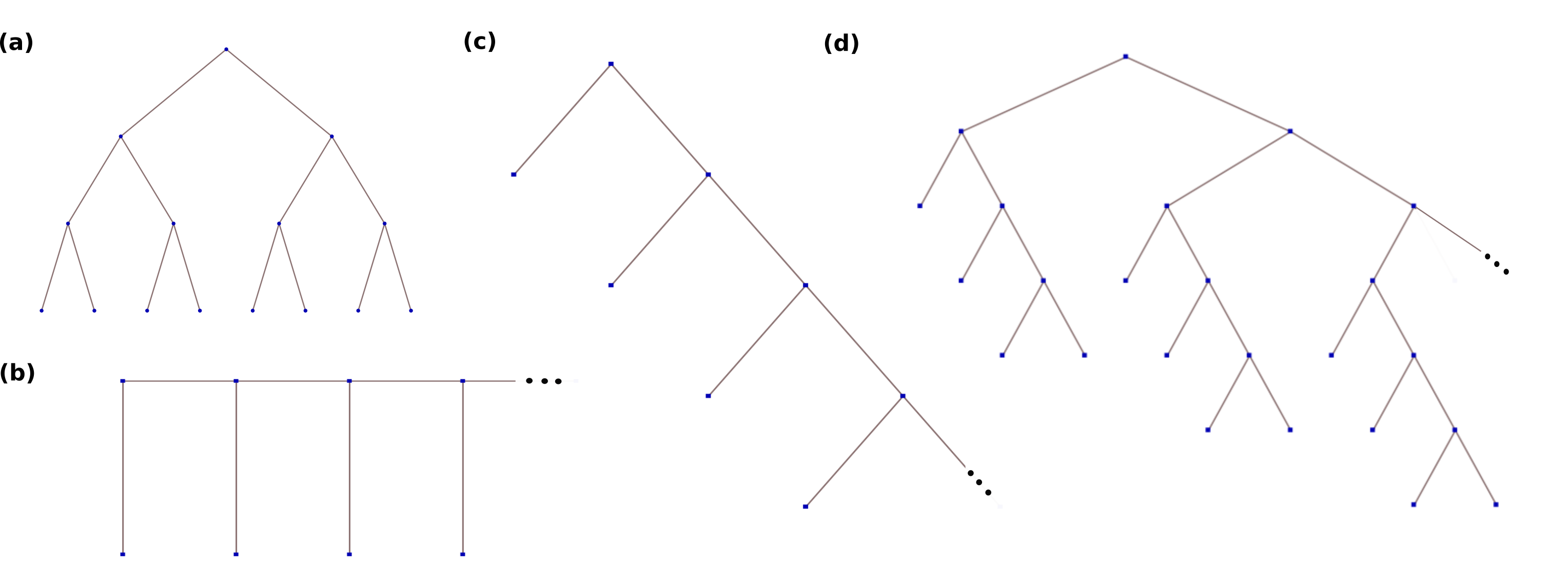}}
\caption{Some examples of Boolean formulas represented as trees.
  Terminal nodes represent logical variables, whereas all other nodes
  represent gates. (a) Full-depth circuit consisting of fan-in~2 gates.
  (b) An example of a shallow circuit of depth~1, for instance
  $a \land b \land c\ldots$, with gates having large fan-in
  (the horizontal line with many nodes represents a single gate in this
  case). (c) Fan-in~2 representation of the
  shallow circuit in (b), for instance
  $a \land (b \land (c\ldots))$. (d) A circuit of depth~2 (such
  as CNFs and DNFs) in fan-in~2 representation.}
\label{fig:trees}
\end{figure*}

Keeping this in mind, we can construct a rigorous approach to convert
any Boolean formula $L$, which contains AND and OR gates, into a DL
circuit $P_L$:
\begin{itemize}
\item Convert the Boolean formula $L$ into a binary tree, that is,
  into a form in which every gate has fan-in~2 (see
  \reffig{fig:trees}). In
  doing so, avoid interpreting configurations of the form
  $\ldots a\land b \land c\ldots$ as $(a\land b)\land c$, and instead
  use $a\land(b\land c)$; the particular configuration
  $(a\land b)\land c$ is not allowed (the reason for this rule will be
  explained later).
\item Select the gate $\lT$ at the root of the tree and construct the
  corresponding DL circuit for $a\lT b$ as shown in
  \reffig{fig:base}.
\item Proceed recursively down the tree, successively following the left and right
  branches. Every such step can be described as
  $a\lT b\to (a \lT_1 b) \lT c$ if we select the left branch, or as
  $a\lT b\to a \lT (b \lT_1 c)$ if we select the right branch, for
  some $\lT$, $\lT_1$. Transfer this step to the DL circuit by
  applying the procedure shown in \reffig{fig:cascading}. In
  particular, when replacing the 2-port elements $P_x$ with a 3-port
  gate, connect the first two outputs in an order-preserving
manner and keep the last one
  ``hanging,'' meaning that at this point a new channel is created.
\end{itemize}

This procedure ensures that at every step we replace the 2-port
devices $P_x$ with 3-port devices exactly as shown in
\reffig{fig:cascading}.

After the DL circuit has been constructed, we run it as follows:
\begin{itemize}
\item Set the operators $P_x$ according to the Boolean variables $x$
  as given in \refeqs{eq:P0}{eq:P1}.
\item Apply the input vector $(1,0,0,\ldots,0)$ to the circuit
  (meaning a nonzero signal in the first channel and zeros in all remaining
  channels), see \reffig{fig:cascading-gen}.
\item Read the output from either the first channel, $c_1$, or the second
channel, $c_2$. The first channel contains $\overline{L}$, that is,
zero if $L=1$ and nonzero if $L=0$, whereas the second channel
contains $L$. All remaining output channels contain zeros.
\end{itemize}

In \reffig{fig:br-and_or_circuit_and_table}, a DL circuit is shown for the
simple formula $L=(a \land b) \lor c$, together with its truth
table, representing the values of all outputs for all values of the
logical variables $a,b,c$. All other DL circuits with 1 and 2 gates
are presented in Appendix~A (\reffigs{fig:a OR (b AND c)}{fig:(a OR b) OR c}). A more complex exemplary circuit for
$L=(((a\land b) \lor c) \land d) \lor (e \land f)$ is given in
\reffig{fig:complex-formulas}.

Close comparison (see Supplementary) shows that the connectivity
topology resulting from this procedure is not fully equivalent to that
of Hardy and Shamir. The differences, however, are presumably confined to paths that are
never traversed.

As we will see in the next subsection, this
property of ``having a $\overline{L}$-output in $c_1$'' and ``having a $L$-output in $c_2$'' is transferred from 1- and 2-gate expressions to
 expressions with an arbitrary number of gates.

\subsection{Universality proof by induction}
\label{subsec:proof-universality}

\begin{figure*}[t]
  \centering

  \includegraphics[width=0.6\linewidth]{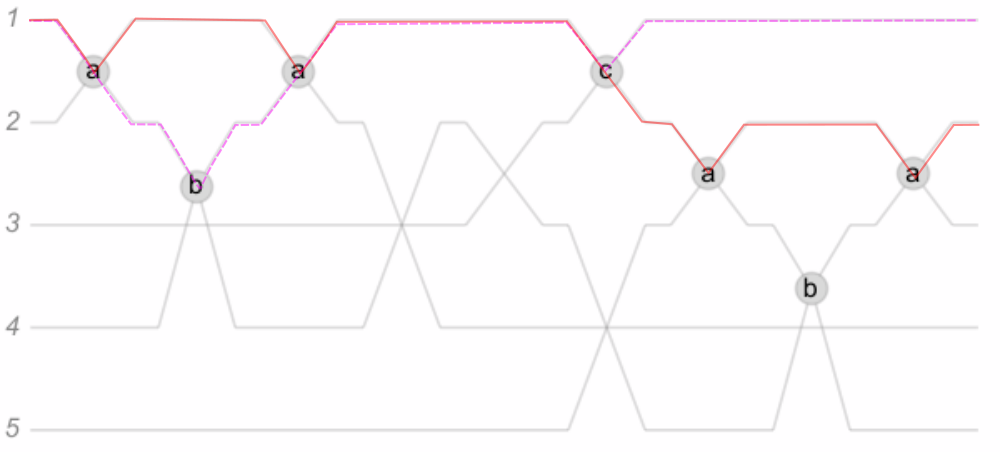}

  \vspace{1em}

  \begin{tabular}{cccc @{\quad} c @{\quad} c @{\quad} ccc}
    \toprule
    $a$ & $b$ & $c$ & $(a\land b)\lor c$ &
      $c_1$ & $c_2$ & $c_3$ & $c_4$ & $c_5$ \\
    \midrule
  0 & 0 & 0 & 0 & 1 & 0 & 0 & 0 & 0 \\
                   0 & 0 & 1 & 1 & 0 & 1 & 0 & 0 & 0 \\
                   0 & 1 & 0 & 0 & 1 & 0 & 0 & 0 & 0 \\
\rowcolor{lightpink}
    0 & 1 & 1 & 1 & 0 & 1 & 0 & 0 & 0 \\
\rowcolor{myviolet}    
                                  1 & 0 & 0 & 0 & 1 & 0 & 0 & 0 & 0 \\
                                                                                                1 & 0 & 1 & 1 & 0 & 1 & 0 & 0 & 0 \\
                                                                                                1 & 1 & 0 & 1 & 0 & 1 & 0 & 0 & 0 \\
                                                                                                1 & 1 & 1 & 1 & 0 & 1 & 0 & 0 & 0 \\
    \midrule
    \multicolumn{4}{l}{$c_i(a,b,c)$:} &
      $\overline{(a\land b)\lor c}$ &
      $(a\land b)\lor c$ &
      $0$ &
      $0$ &
      $0$ \\
    \bottomrule
  \end{tabular}
  \caption{Exemplary DL circuit representing $L=(a\land b)\lor c$
    (upper panel) and the corresponding truth table (lower panel) for
    each of the outputs $c_i(a,b,c)$.  DL is made as a composition of two gates
    $x=a\land b$ and $x\lor c$ prepared according to the rule
    \reffig{fig:cascading}c,d; the rectangles show sub-gates $a\land b$.  For the graphical
    representation, we use compact notation for the optical elements.
    Namely, every circle labeled with $x=a,b,\ldots$ represents the
    element $P_x$ according to \refeqs{eq:P0}{eq:P1}. Note that
    $P_x^{-1}=P_x$, so both direct and iverse elements are represented
    by the same symbols. The colored lines represent the light paths
    for two exemplary cases: $a=0$, $b=1$, $c=1$ (solid pale pink
    line), corresponding to $L=1$, and $a=1$, $b=0$, $c=0$ (magenta
    dashed line), corresponding to $L=0$. For convenience, the table
    also contains, together with the values of $a,b,c$, the values of
    the function $L(a,b,c)$. The lowest row of the table summarizes
    the outputs $c_i(a,b,c)$ in the form of Boolean functions. The
    cases corresponding to the exemplary light paths in the circuit
    above are highlighted using the corresponding colors.}
  \label{fig:br-and_or_circuit_and_table}
\end{figure*}

\begin{table*}[t]
  \centering
  \begin{tabular}{l @{\qquad}  l @{\qquad} l @{\qquad} l @{\qquad} l @{\quad} l}
    \toprule
    Boolean formula & $c_1$ & $c_2$ & $c_3$ & $c_4$ & $c_5$ \\
    \midrule

    $a\land b$
      & $\overline{a\land b}$ & $a\land b$ & $0$ &  &  \\
    $a\lor b$
      & $\overline{a\lor b}$ & $a\lor b$ & $0$ &  &  \\
    $a\lor (b\land c)$
      & $\overline{a\lor (b\land c)}$ & $a\lor (b\land c)$ & $0$
      & $0$ &  \\
    $(a\lor b)\land c$
      & $\overline{(a\lor b)\land c}$ & $(a\lor b)\land c$ & $0$
      & $0$ & $0$ \\
    $a\land (b\lor c)$
      & $\overline{a\land (b\lor c)}$ & $a\land (b\lor c)$ & $0$
      & $0$ &  \\
    $(a\land b)\lor c$
      & $\overline{(a\land b)\lor c}$ & $(a\land b)\lor c$ & $0$
      & $0$ & $0$ \\
    $a\land (b\land c)$
      & $\overline{a\land (b\land c)}$ & $a\land (b\land c)$ & $0$
      & $0$ &  \\
$(a\land b)\land c$
& $\overline{(a\land b)\land c}$ & $(a\land b)\land c$ & $0$
& $0$ & $0$ \\
    $a\lor (b\lor c)$
      & $\overline{a\lor (b\lor c)}$ & $a\lor (b\lor c)$ & $0$
      & $0$ &  \\
    $(a\lor b)\lor c$
      & $\overline{(a\lor b)\lor c}$ & $(a\lor b)\lor c$ & $0$
      & $0$ & $0$ \\
    \bottomrule
  \end{tabular}
\caption{Truth tables for the outputs $c_i$ of the basic direct
  circuits. The truth tables are given in the condensed form of
  Boolean functions producing the same outputs for all values of
  $a$, $b$, and $c$.  The
  complete truth tables, as well as graphical representations of the
  corresponding circuits, can be found in Appendix~A (\reffigs{fig:and_or}{fig:(a OR b) OR c}) and
  \reffig{fig:br-and_or_circuit_and_table}. Note nonassociativity of
  DL. It can be seen that
  $c_1$ always contains the negation of the corresponding Boolean
  formula, whereas $c_2$ contains the formula itself. All other
  outputs contain zeros.}
  \label{tab:summary_ci}
\end{table*}

The procedure described above suggests that for any Boolean circuit
$L$ we can construct an equivalent DL circuit $P_L$, which calculates
$\overline{L}$ and $L$ at the output of the first two channels $c_1$, $c_2$. For small
circuits this can be tested directly, which we did for all 1- and
2-gate circuits (see Table~\ref{tab:summary_ci} and
\reffigs{fig:and_or}{fig:(a OR b) OR c} in Appendix~A, as well as \reffig{fig:br-and_or_circuit_and_table}). We cannot make such
a direct test for an arbitrary number of gates; therefore, a proof is
necessary. We prove this statement by mathematical induction: assuming that for step $n$ (in the procedure described
above) the property we want to prove (the induction hypothesis) holds, we
prove it for step $n+1$. For $n=1$, the proof is provided by
direct verification.

Before we start, let us formulate more carefully the property we want to
prove. We observe that there are in fact two separate independent
properties to be proved: (i) We can map every logical circuit $L$ into
a DL circuit which evaluates $L$, and (ii) We can map the logical
refinement step (such as $a\lT_1 b\to (a \lT_2 b) \lT_1 c$ for some
gates $\lT_i$, $i=\{1,2\}$) into the corresponding refinement step in
the DL circuit. 
%
 With this in mind we formulate i) and ii) more formally, as two
 induction hypotheses:

\medskip
\noindent
\textbf{(H1) Circuit mapping.}
For every Boolean expression
\[
L(a_1,a_2,\dots,a_m),
\]
constructed from the variables \(a_j\) using the operations
\(\land,\lor\), the nesting algorithm described above produces a DL-circuit
\[P_{L(a_1,\dots,a_m)}\]
such that, for any assignment of the electronic
controls \(a_j\in\{0,1\}\), the first optical  output of the circuit evaluates
the Boolean function \(\overline{L}\) whereas the second optical  output of the circuit evaluates
the Boolean function \(L\), and all other outputs contain zeros. 

\medskip
\noindent
\textbf{(H2) Refinement mapping.} Let \(L(a_1,\dots,a_k,\dots,a_m)\)
be an arbitrary Boolean expression, and let one of its variables
\(a_k\) be replaced by a depth-1 Boolean subexpression,
\[
a_k \;\mapsto\; c_1 \, T \, c_2,
\qquad T\in\{\land,\lor\}.
\]
Then the corresponding DL circuit undergoes the transformation
\[
P_{L(a_1,\dots,a_k,\dots,a_m)}
\;\longrightarrow\;
P_{L(a_1,\dots,(c_1Tc_2),\dots,a_m)},
\]
by replacing the physical 2-port device \(P_{a_k}\) (and the
corresponding inverse \(P_{a_k}^{-1}\)) with the 3-port DL realization
of the gate \(c_1Tc_2\), according to the nesting rules described
above. The resulting circuit evaluates the refined Boolean expression.

\begin{figure*}[htpb] 
  \center{\includegraphics[width=0.8\textwidth]{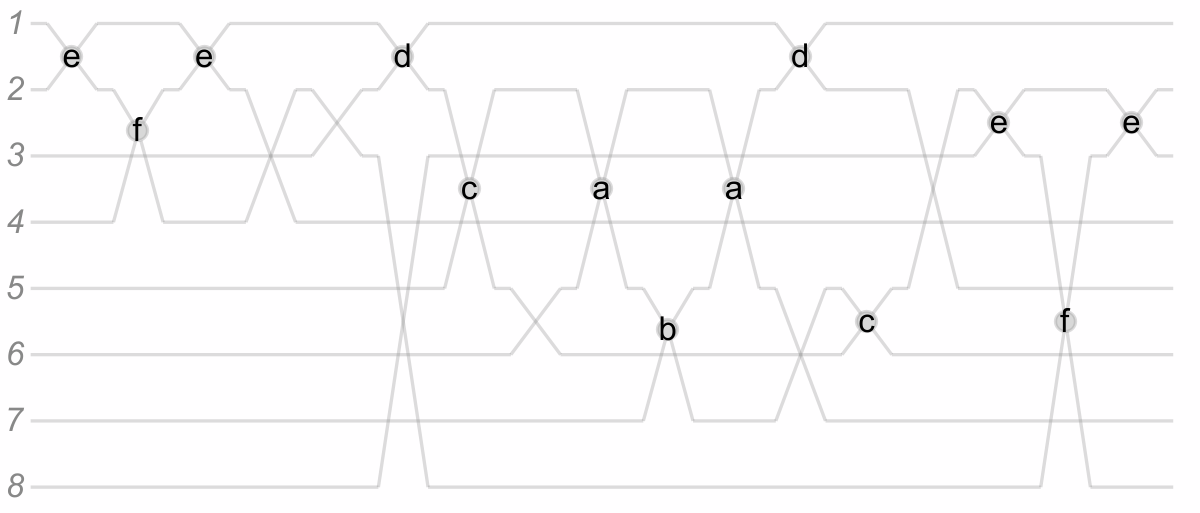}} 
  \caption{An exemplary DL circuit for
    $L=(((a\land b) \lor c) \land d) \lor (e \land f)$ realized in the
    optimal form with the right-branching (~\reffig{fig:cascading}a,b)
    was used whenever possible [$L=((e \land f) \lor (d \land (c
    \lor (a \land b))))$]. According to
    the general procedure described above, the input is $c_1=1$,
    $c_i=0$ for $i>1$. The main output channels are $c_1$ and $c_2$,
    containing $\overline{L}$ and $L$, respectively. All other
    output channels contain zeros. }
\label{fig:complex-formulas}
\end{figure*}

After formulating the induction hypotheses (H1) and (H2), we can
continue with their proof. First, we verify (H1) and (H2) directly for
\(n=1\), i.e., for the simplest possible case, in order to establish
the basis for the induction. Specifically, we construct a single gate
[step (H1)] and extend it to all possible two-gate configurations
[step (H2)]. In other words, proving the case \(n=1\) requires
checking all possible single- and two-gate configurations. This is
done in Table~\ref{tab:summary_ci}, which lists, for every channel,
the corresponding Boolean expression, and in Appendix~A (\reffigs{fig:and_or}{fig:(a OR b) OR c}), where the
corresponding circuits and the values of all output channels are given
explicitly. Table~\ref{tab:summary_ci} verifies that in all the
simplest cases, (H1) and (H2) hold. 

Having verified the case \(n=1\), we proceed by induction. We assume
that (H1) and (H2) hold for all Boolean expressions whose nesting
depth does not exceed \(n\). Let \(L(a_1,\dots,a_m)\) be such
an expression. Consider a refinement of the expression obtained by
replacing an arbitrary variable \(a_k\) inside \(L\) by a binary
subexpression \(c_1Tc_2\). By (H1), the DL circuit
\(P_{L(a_1,\dots,a_m)}\) correctly evaluates \(L\). Applying a single
nesting operation---namely, replacing the controlled 2-port device
\(P_{a_k}\) and its inverse by the 3-port DL realization of
\(c_1Tc_2\)---produces a new circuit. By construction, this circuit is
exactly \(P_{L(a_1,\dots,(c_1Tc_2),\dots,a_m)}\) and, by (H2),
correctly evaluates the refined Boolean expression. Thus, both
hypotheses remain valid for expressions of nesting depth \(n+1\),
thereby completing the proof.

\section{Scalability of the procedure}
\label{sec:scaling}

As shown in the previous sections, nested cascading enables the
implementation of any Boolean formula (represented as a tree) using DL
circuits. At this point, it is important to distinguish between
Boolean formulas and the more general concept of Boolean circuits.
Some examples of Boolean trees are shown in \reffig{fig:trees}. In a
Boolean tree, every node has exactly one parent, and every
intermediate result is therefore used only once. This is precisely the
property that distinguishes Boolean formulas from general Boolean
circuits. By contrast, a Boolean circuit is a directed acyclic graph,
in which intermediate results may be shared and reused by multiple
subsequent gates. General Boolean circuits cannot be realized by
cascading DL elements alone, since reusing an intermediate result
requires converting the corresponding optical signal back into an
electronic representation before it can control subsequent DL gates.
The natural question is then how efficiently Boolean formulas can be realized using nested DL circuits. More precisely, we investigate how the number of optical elements, $N_\mathrm{opt}$, scales with the number of logical gates, 
$N_\mathrm{bool}$, for several practically important classes of Boolean formulas.

The basic gates in \reffig{fig:base} contain two operators $P$ for the
first argument $a$. This leads to a risk that at every step of
expansion the size of the device will double (see \reffig{fig:DL}d),
leading to an exponential growth of $N_\mathrm{opt}$ with
$N_\mathrm{bool}$. This happens in particular if at every expansion
step we add new gates into this ``bad'' slot $a$. Can we avoid this by
a clever arrangement of the Boolean gates?

This is indeed always the case for shallow circuits of depth~1 such as
those presented in \reffig{fig:trees}b. When reformulating such
circuits into fan-in~2 form, as required by our procedure
(\reffig{fig:trees}c), we can always put the continuation of the
circuit into the ``good,'' non-duplicating slot $b$, whereas the
``bad,'' duplicating slot appears only in the
terminal nodes (variables). That is, the Boolean expression
$a \land b \land c\ldots$ must be converted into the fan-in~2 form as
$a \land (b \land (c\ldots))$ rather than as
$((\ldots(a \land b) \land c)\ldots$ (and similarly for $\lor$; for
$\land$ chains, in addition, the configuration $(a\land b) \land c$
is additionally prohibited, as discussed above). For the depth-1 circuit shown in
\reffig{fig:trees}b, we therefore obtain only a constant overhead,

\begin{equation}
  \label{eq:scaling_depth_1}
  N_\mathrm{opt} \sim N_\mathrm{bool}.
\end{equation}

Will we observe something similar in more complex cases? Let us first
focus on a few representative classes of Boolean expressions:
conjunctive and disjunctive normal forms (CNFs and DNFs) (see
\reffig{fig:trees}d), fully nested binary trees
(\reffig{fig:trees}a) with alternating logical operations, and binary
adders. Importantly, in the scalings
$N_\mathrm{opt}(N_\mathrm{bool})$ below we compare the DL circuits to
the corresponding Boolean formulas (which were used to obtain these DL
circuits) rather than to more general and potentially more optimal
Boolean circuits.

\subsection{Conjunctive and disjunctive normal forms}

A CNF is a Boolean expression of the form
\begin{equation}
  \label{eq:cnf}
  L_{\mathrm{CNF}} =
  \bigwedge_{i=1}^{m}
  \left(
    \bigvee_{j=1}^{k_i} x_{ij}
  \right),
\end{equation}
where $x_{ij}$ are Boolean variables (possibly negated). After
converting all logical operations to fan-in--2 form (cf.
\reffig{fig:trees}d), the number of Boolean gates scales as
$N_{\mathrm{bool}} \sim \sum_i k_i$.

For CNFs, the nested DL construction exhibits particularly favorable
scaling, as follows from \reffig{fig:trees}d. Indeed, each
OR clause by itself is a depth-1 circuit and demonstrates only linear
scaling according to \refeq{eq:scaling_depth_1}. As we proceed down
the tree in \reffig{fig:trees}d, we must add every OR subtree into the
doubling slot only once. That is, effectively, our procedure is
equivalent to doubling the OR subtrees, which leads to only a
constant overhead, preserving the linear scaling for this case:
\begin{equation}
  \label{eq:cdf-scal}
  N_\mathrm{opt} \sim N_\mathrm{bool}.  
\end{equation}

And indeed, this reasoning is fully supported by numerical
simulations. In our numerical simulations, we calculated
$N_\mathrm{opt}(N_\mathrm{bool})$ using random CNF forms of
increasing size, controlled by the increasing parameter $n$ with
$k_i=n$ and $m=n$ in \refeq{eq:cnf}, for
$n$ up to $n_\mathrm{max}=20$. We indeed
found the linear scaling
\begin{equation}
  N_{\mathrm{opt}} \approx -20.80 + 5.69\, N_{\mathrm{bool}}
\end{equation}
as given by the fit (see \reffig{fig:scaling}). The reasoning above
confirms that shallow Boolean circuits with a fixed small depth can be
implemented efficiently within DL. Exactly the same reasoning is also
applicable to DNFs (which have the same structure as CNFs).

\subsection{Fully nested binary trees}

As a further important example, we consider fully nested Boolean
expressions (see \reffig{fig:trees}a) with alternating logical
operations, represented as complete binary trees, for instance
\begin{equation}
  L_{\mathrm{tree}} =
  \bigl(
    (x_{11} \lor x_{12})
    \land
    (x_{21} \lor x_{22})
  \bigr)
  \lor
  \bigl(
    (x_{31} \lor x_{32})
    \land
    (x_{41} \lor x_{42})
  \bigr),
\end{equation}
and recursively extended to larger depths.

\begin{figure*}[htpb]
  \centering
  \includegraphics[width=1\linewidth]{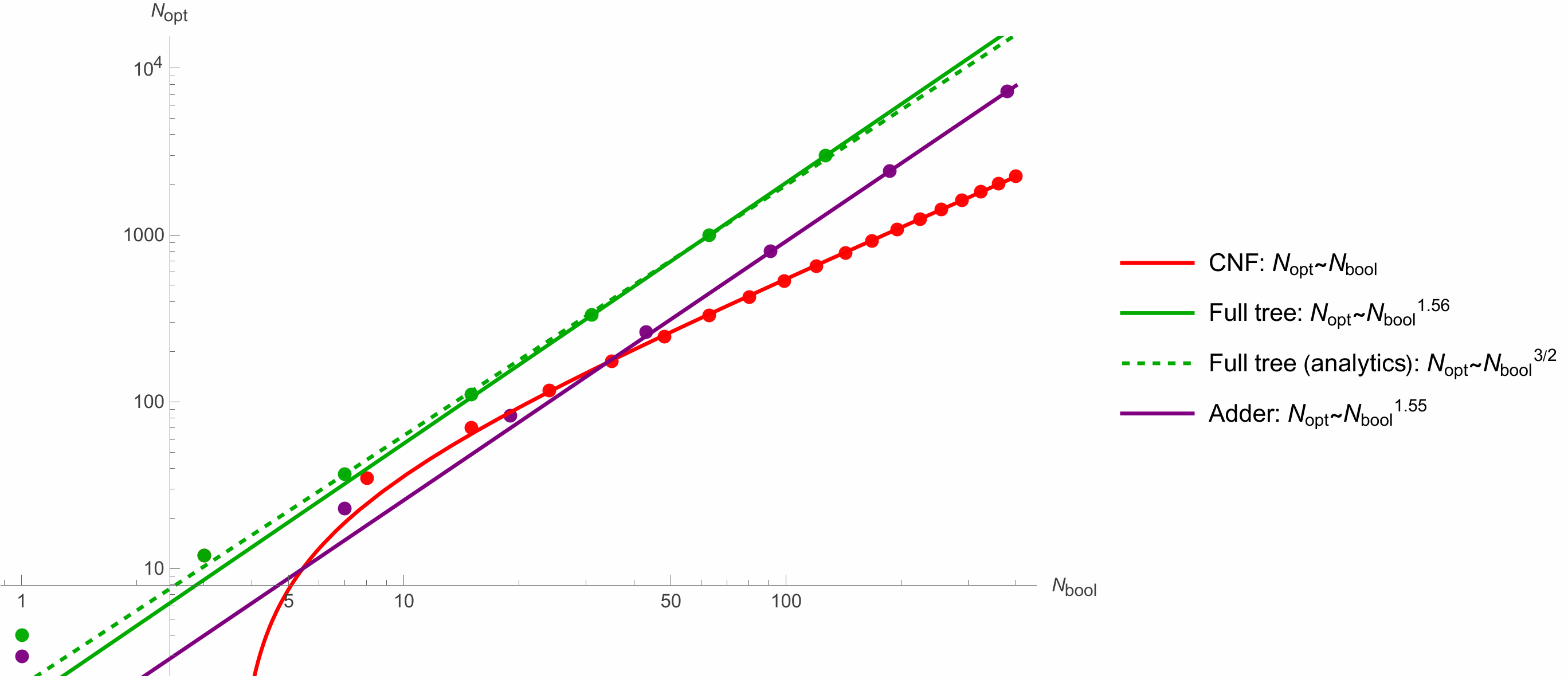}
\caption{Scaling of the optical gate count $N_{\mathrm{opt}}$ as a
  function of the Boolean gate count $N_{\mathrm{bool}}$ for
  different classes of Boolean formulas realized using nested
  directed logic. Symbols show numerically calculated data points,
  whereas the solid lines represent fits to these data. The dashed
  line shows the analytically obtained ``most typical'' power-law
  scaling
  $N_{\mathrm{opt}} \sim N_{\mathrm{bool}}^{3/2}$. Red
  symbols show conjunctive normal forms (CNFs), green symbols
  correspond to a full binary tree, whereas purple symbols represent
  the carry-out bit of an $n$-bit ripple-carry adder.}
\label{fig:scaling}
\end{figure*}

In such expressions, both inputs of a logical gate are themselves
nontrivial subexpressions. This is the most unfortunate situation for
the nested DL construction, since it implies that at each level, one
of the subcircuits comes into the doubling slot of a DL gate and must
inevitably be duplicated. Thinking naively, this should lead to an
exponential scaling of $N_{\mathrm{opt}}$ with $N_{\mathrm{bool}}$,
but this does not happen. Numerically, we find a power-law dependence
\begin{equation}
  \label{eq:Nopt-vs-Nbool-tree-num}
  N_{\mathrm{opt}} \approx a\, N_{\mathrm{bool}}^{k},
  \qquad
  a \approx 1.55,\quad k \approx 1.56,
\end{equation}
see \reffig{fig:scaling}, where we considered tree depths
up to $n=7$.

This power-law scaling can be understood analytically. For this, let us
consider a depth-2 tree. Without duplication, the number of elements
on the lower level would be twice as large (by the very nature of a
binary tree). With the duplication of one of the slots, the number of
elements becomes three times larger instead. Therefore, we conclude that
\begin{equation}
  \label{eq:Nopt-tree}
  N_\mathrm{opt}\sim 3^n,
\end{equation}
whereas for the Boolean gates we have obviously
\begin{equation}
  \label{eq:Nbool-tree}
  N_\mathrm{bool}\sim 2^n.
\end{equation}
Expressing $N_\mathrm{opt}$ as a function of $N_\mathrm{bool}$ from
\refeqs{eq:Nopt-tree}{eq:Nbool-tree}, we obtain
\begin{equation}
  \label{eq:Nopt-vs-Nbool-tree}
  N_\mathrm{opt}\sim N_\mathrm{bool}^{3/2}. 
\end{equation}
This analytical scaling is quite close to the one obtained numerically
(see \reffig{fig:scaling} and
\refeq{eq:Nopt-vs-Nbool-tree-num}), where the dashed line represents
the power law
$N_\mathrm{opt}=2 N_\mathrm{bool}^{3/2}$.

\subsection{Adder circuit}

In this subsection we consider the Boolean realization of an $n$-bit
ripple-carry adder, which computes the sum of two binary numbers
$(a_1,\ldots,a_n)$ and $(b_1,\ldots,b_n)$, where $a_1$ and $b_1$
denote the least significant bits. The adder produces a sequence of
sum bits $(s_1,\ldots,s_n)$ and a final carry-out bit $c_{n+1}$. The
computation proceeds sequentially from the least significant to the
most significant bit, with each bit depending on the corresponding
input bits and the incoming carry.

For each bit position $i$, the sum bit is given by
\begin{equation}
  s_i = a_i \oplus b_i \oplus c_i,
\end{equation}
where  $c_i$ is the carry into
the $i$-th bit. The initial carry is set to $c_1 = 0$. The carry bits
are then generated recursively according to
\begin{equation}
  \label{eq:carry-bit}
  c_{i+1} =
  (a_i \land b_i)
  \lor
  (a_i \land c_i)
  \lor
  (b_i \land c_i),
  \qquad
  c_1 = 0.
\end{equation}
Here we focus on the carry-out bit $c_{n+1}$.

In this case, the numerically observed scaling (for $n\le 7$) again
follows a power law,
\begin{equation}
  \label{eq:eq:Nopt-vs-Nbool-adder}
  N_{\mathrm{opt}} \approx a\, N_{\mathrm{bool}}^{k},
  \qquad
  a \approx 0.73,\quad k \approx 1.55.
\end{equation}
The exponent is close to that obtained for the fully nested tree,
whereas the prefactor $a$ is smaller, so that
$N_{\mathrm{opt}}$ for the adder is strictly smaller than
$N_{\mathrm{opt}}$ for the full binary tree. This is not accidental, since the tree representing the adder is
obviously a subtree of the full binary tree.  This relation is discussed in more detail in the next subsection.

It is important to emphasize that if the adder were implemented as a
Boolean circuit, allowing intermediate results to be shared and reused
in recursive expressions such as \refeq{eq:carry-bit}, it would
exhibit linear scaling, $N_{\mathrm{bool}}\sim n$. We, however,
consider $N_{\mathrm{bool}}$ for the corresponding Boolean formula (rather than a circuit); when treating
$c_i$ as a Boolean formula, the carry recursion in
\refeq{eq:carry-bit} involves repeated substitution of $c_i$ into
$c_{i+1}$, which leads to exponential growth of $N_{\mathrm{bool}}$
with $n$. For DL circuits, sharing and reuse are also not available. That is, for the adder in
\refeq{eq:eq:Nopt-vs-Nbool-adder}, both $N_{\mathrm{bool}}$ and
$N_{\mathrm{opt}}$ are exponential in $n$,  yet $N_{\mathrm{opt}}$
grows only polynomially with $N_{\mathrm{bool}}$ as given by \refeq{eq:eq:Nopt-vs-Nbool-adder}. 

\subsection{Scalings for worst and most typical cases}
\label{sec:general-most-typical}

Returning to the question of what the ``worst,'' ``general,'' or
``most typical'' scaling \(N_{\mathrm{opt}}(N_{\mathrm{bool}})\) is,
we notice that, in fact, the full binary tree represents the most
typical type of Boolean formula. Indeed, most Boolean formulas are
full binary trees (cf. \reffig{fig:trees}a) with some arrangement
of gates at the nodes. The remaining Boolean formulas are binary
trees with some of the edges removed. Therefore, the power-law scaling
\refeq{eq:Nopt-vs-Nbool-tree} must be considered very close to the
worst-case scaling and, at the same time, as the most typical
scaling. The case of the adder considered above is a very good
illustration of this point: the adder is obviously ``smaller'' than
the full tree, and therefore its scaling must be milder, which we also
readily observe in \reffig{fig:scaling}.

Note, however, that this scaling, as already mentioned, represents the
relation between the DL circuit size and the size of the corresponding
Boolean formula, which, in turn, may exhibit scaling that is
exponentially worse than that of a Boolean circuit, which
allows sharing and reuse of intermediate results, as in the
case of the adder discussed above. On the other hand, as we see in
\reffig{fig:complex-formulas}, quite a large part of the circuit can be
removed since it does not influence the value in the ``main'' channel
$c_1$. Consideration of the influence of such optimizations on the
scaling is, however, beyond the scope of this paper.

\section{Conclusions and Discussion}
\label{sec:conclusions}

In this work, we have revisited the nested cascading in directed
logic, proposed by Hardy and Shamir in \cite{hardy07}. Despite its
great promise, the ingenious idea of Hardy and Shamir has been up to
now neither properly proved nor thoroughly tested beyond some of the few-gates
configurations.

We filled these gaps by constructing an unambiguous cascading
procedure and providing a rigorous proof of its correctness. We showed
that the main promise of this approach does indeed hold: for every
Boolean formula \(L\), consisting of AND and OR operations and
converted into fan-in-2 form, one can construct, under certain
composition rules, a DL circuit that evaluates both \(L\) and
\(\overline{L}\). In this article, we focus only on AND and OR gates,
since, once the negated variables are available, these gates alone are
sufficient to represent any Boolean formula.

Our proof is based on mathematical induction. By its very nature, an
inductive proof derives the validity of arbitrarily complex gates from
the validity of elementary gates and their simplest compositions,
provided that the latter are verified directly. This direct verification
included 72 different configurations, all of which were tested by us.

In general, for purely optical DL circuits, without back-conversion of
optical signals into electronic ones, sharing and reuse of intermediate
results are not possible. Boolean formulas, in contrast to the more
general Boolean circuits, also do not allow such sharing. As a
consequence, the size of a Boolean formula can, in some cases, be
exponentially larger than that of the corresponding optimal Boolean
circuit.

On the other hand, even when comparing the nested DL implementation with
the corresponding Boolean formula, one might expect a further exponential
increase in the number of DL elements, since the nested DL implementation
requires one of the arguments to be duplicated. We analyzed this question
and showed that both the worst-case and the typical scaling of the number
of DL elements relative to the corresponding Boolean formula are not
exponential, but are asymptotically given by a power law with an exponent
close to (3/2). Similar scaling is observed for important algorithms
such as addition.

Thus, the nested DL construction faithfully reflects the complexity of
Boolean formulas, or trees, rather than that of Boolean circuits. The
size of DL circuits is generally at most polynomially larger than that of
the corresponding Boolean formula, while both can be exponentially larger
than the corresponding optimal Boolean circuit. However, there are
important cases, such as conjunctive and disjunctive normal forms, in
which the DL circuit scales linearly with both the corresponding Boolean
formula and the corresponding Boolean circuit.

In general, the difficulty of DL logic in comparison with optimal
Boolean circuits is not the duplication of variables, but rather the lack
of sharing and reuse of intermediate results. Our results also indicate
that the points at which such sharing and reuse become necessary are the
most natural places to go beyond pure DL cascading and use mixed
architectures, including back-conversion from optical to electronic
signals, optical nonlinearities, or other nonlinear mechanisms.


\clearpage

\appendix
\onecolumngrid

\section{Appendix A: truth tables and graphical representations of 1-
  and 2-operator Boolean formulas}

\begin{figure*}[hp]
  \centering

  \begin{minipage}[hp]{0.48\textwidth}
    \centering

    \includegraphics[width=0.9\linewidth]{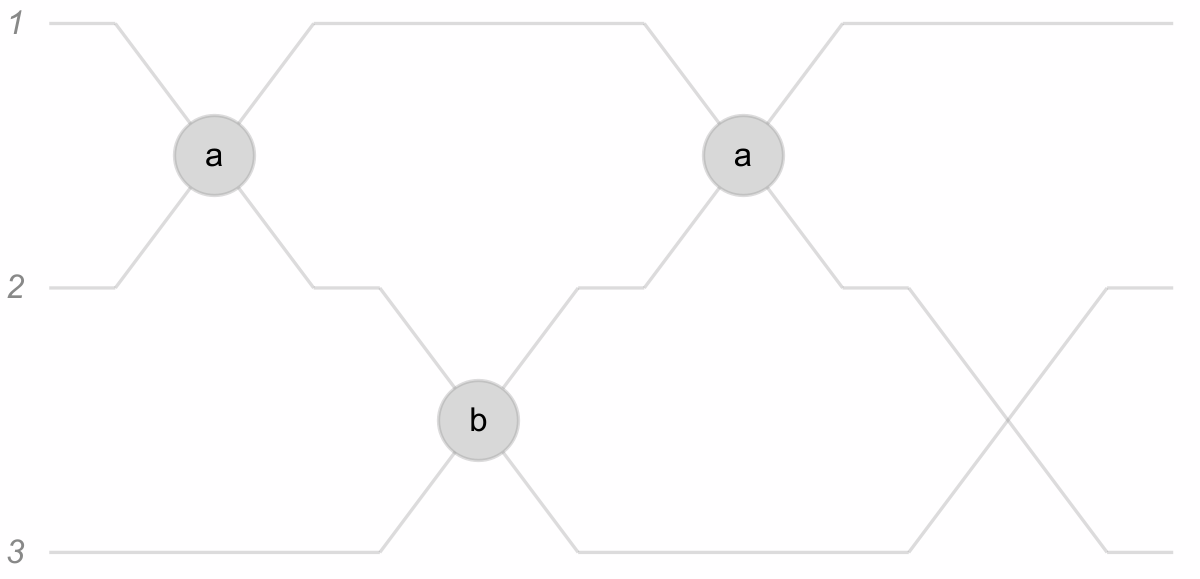}

    \vspace{0.8em}

    \begin{tabular}{ccc @{\quad} ccc}
      \toprule
      $a$ & $b$ & $a\land b$ & $c_1$ & $c_2$ & $c_3$ \\
      \midrule
      0 & 0 & 0 & 1 & 0 & 0 \\
      0 & 1 & 0 & 1 & 0 & 0 \\
      1 & 0 & 0 & 1 & 0 & 0 \\
      1 & 1 & 1 & 0 & 0 & 1 \\
      \midrule
      \multicolumn{3}{l}{$c_i(a,b)$:} &
        $\overline{a\land b}$ & $0$ & $a\land b$ \\
      \bottomrule
    \end{tabular}

    \vspace{0.3em}
    
    {\small (a) $a\land b$}
  \end{minipage}
  \hfill
  \begin{minipage}[hp]{0.48\textwidth}
    \centering

    \includegraphics[width=0.9\linewidth]{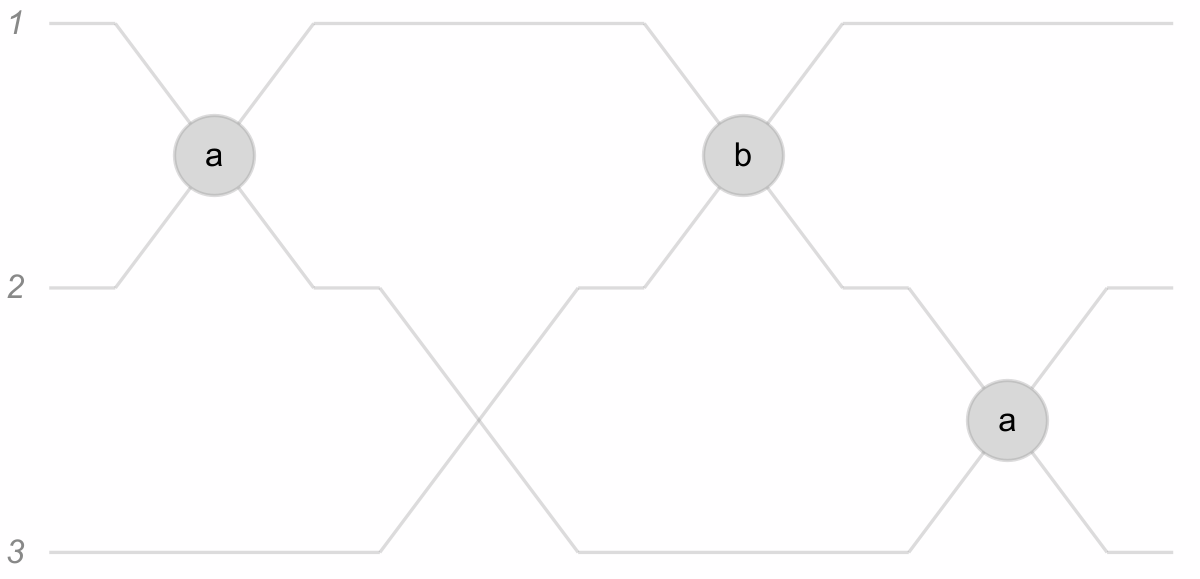}

    \vspace{0.8em}

    \begin{tabular}{ccc @{\quad} ccc}
      \toprule
      $a$ & $b$ & $a\lor b$ & $c_1$ & $c_2$ & $c_3$ \\
      \midrule
      0 & 0 & 0 & 1 & 0 & 0 \\
      0 & 1 & 1 & 0 & 1 & 0 \\
      1 & 0 & 1 & 0 & 1 & 0 \\
      1 & 1 & 1 & 0 & 1 & 0 \\
      \midrule
      \multicolumn{3}{l}{$c_i(a,b)$:} &
        $\overline{a\lor b}$ & $a\lor b$ & $0$ \\
      \bottomrule
    \end{tabular}

    \vspace{0.3em}
    
    {\small (b) $a\lor b$}
  \end{minipage}
\caption{DL circuits for binary logic operations. (a) $a\land b$
  and (b) $a\lor b$, showing the graphical realizations (top) and the
  corresponding outputs $c_i(a,b)$ (bottom). The figure is
  equivalent to \reffig{fig:base}, taking into account that
  \reffig{fig:base} presents the generally scalable configuration,
  whereas here its instantiation is given in the form of a 1-gate
  circuit. 
}
  \label{fig:and_or}
\end{figure*}

In this appendix, the truth tables and the graphical representation of
all two-variable expressions $a \,\mathrm{T}\, b$,
$\mathrm{T}\in\{\land,\lor\}$ as well as thee-variable expressions $(a
\,\mathrm{T_1}\, b) \, \mathrm{T_2}\, c$ and $a
\,\mathrm{T_1}\, (b \, \mathrm{T_2} \,c)$, 
$\mathrm{T}_i\in\{\land,\lor\}$, $i=1,2$ in \reffigs{fig:and_or}{fig:(a OR b) OR c}.   Note that the circuit and the truth table for the case
$(a\land b)\lor c$ are already shown in
\reffig{fig:br-and_or_circuit_and_table} and therefore are not given
in this Appendix. Denotations for all the figures here are as in \reffig{fig:br-and_or_circuit_and_table}.


\begin{figure*}[hp]
  \centering

  \includegraphics[width=0.5\linewidth]{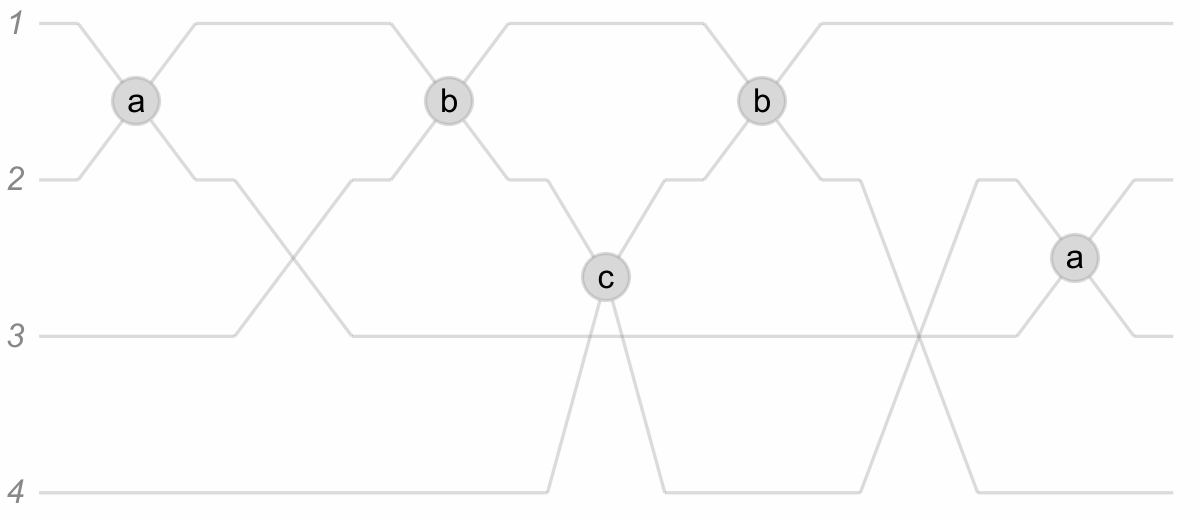}

  \vspace{1em}

  \begin{tabular}{cccc @{\quad} ccccc}
    \toprule
    $a$ & $b$ & $c$ & $a\lor (b \land c)$ &
      $c_1$ & $c_2$ & $c_3$ & $c_4$ \\
    \midrule
0 & 0 & 0 & 0 & 1 & 0 & 0 & 0 \\
0 & 0 & 1 & 0 & 1 & 0 & 0 & 0 \\
0 & 1 & 0 & 0 & 1 & 0 & 0 & 0 \\
0 & 1 & 1 & 1 & 0 & 1 & 0 & 0 \\
1 & 0 & 0 & 1 & 0 & 1 & 0 & 0 \\
1 & 0 & 1 & 1 & 0 & 1 & 0 & 0 \\
1 & 1 & 0 & 1 & 0 & 1 & 0 & 0 \\
1 & 1 & 1 & 1 & 0 & 1 & 0 & 0 \\
    \midrule
    \multicolumn{4}{l}{$c_i(a,b,c)$:} &
      $\overline{a\lor (b \land c)}$ &
      $a\lor (b \land c)$ &
      $0$ &
      $0$ \\
    \bottomrule
  \end{tabular}

  \caption{Direct circuit for $a \lor (b \land c)$: (top) graphical
  realization and (bottom) the corresponding outputs $c_i(a,b,c)$. }
  \label{fig:a OR (b AND c)}
\end{figure*}

\begin{figure*}[h]
  \centering

  \includegraphics[width=0.55\linewidth]{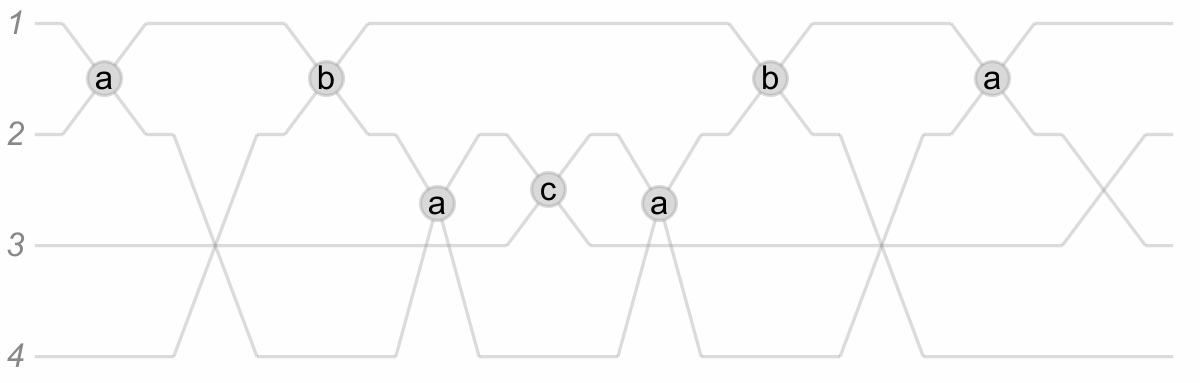}

  \vspace{1em}

  \begin{tabular}{cccc @{\quad} ccccc}
    \toprule
    $a$ & $b$ & $c$ & $(a\lor b)\land c$ &
      $c_1$ & $c_2$ & $c_3$ & $c_4$ & $c_5$ \\
    \midrule
    0 & 0 & 0 & 0 & 1 & 0 & 0 & 0 & 0 \\
    0 & 0 & 1 & 0 & 1 & 0 & 0 & 0 & 0 \\
    0 & 1 & 0 & 0 & 1 & 0 & 0 & 0 & 0 \\
    0 & 1 & 1 & 1 & 0 & 1 & 0 & 0 & 0 \\
    1 & 0 & 0 & 0 & 1 & 0 & 0 & 0 & 0 \\
    1 & 0 & 1 & 1 & 0 & 1 & 0 & 0 & 0 \\
    1 & 1 & 0 & 0 & 1 & 0 & 0 & 0 & 0 \\
    1 & 1 & 1 & 1 & 0 & 1 & 0 & 0 & 0 \\
    \midrule
    \multicolumn{4}{l}{$c_i(a,b,c)$:} &
      $\overline{(a\lor b)\land c}$ &
      $(a\lor b)\land c$ & $0$ &
      $0$ &
      $0$ \\
    \bottomrule
  \end{tabular}

  \caption{Direct circuit for $(a\lor b)\land c$ and the corresponding outputs
  $c_i(a,b,c)$.}
  \label{fig:(a OR b) AND c}
\end{figure*}

\begin{figure*}[t]
  \centering

  \includegraphics[width=0.5\linewidth]{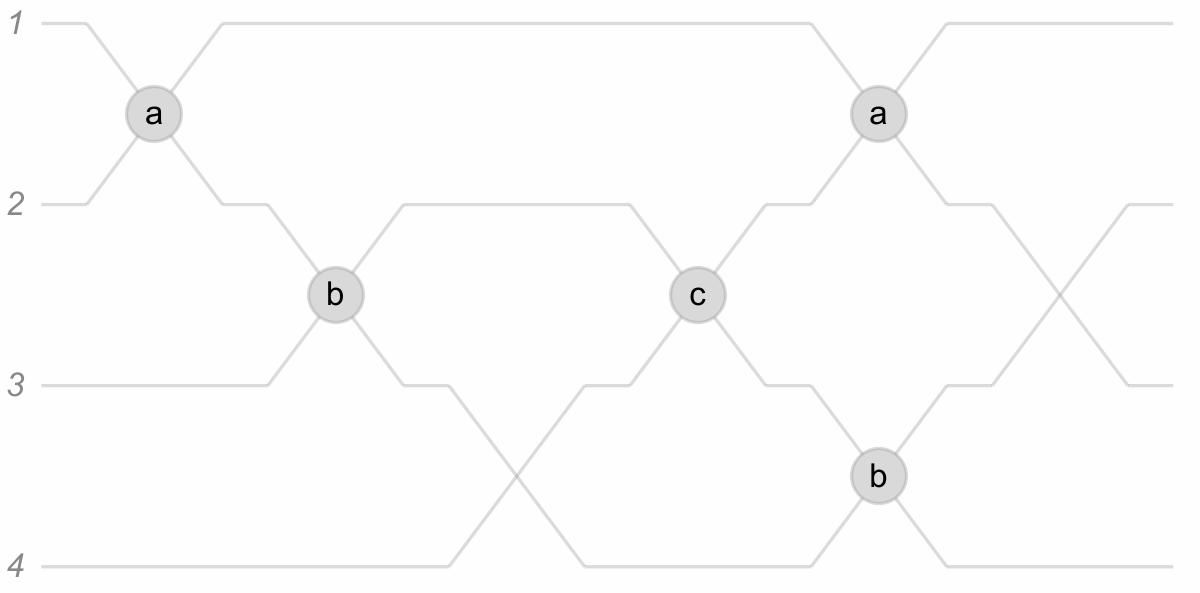}

  \vspace{1em}

  \begin{tabular}{cccc @{\quad} cccc}
    \toprule
    $a$ & $b$ & $c$ & $a\land (b \lor c)$ &
      $c_1$ & $c_2$ & $c_3$ & $c_4$ \\
    \midrule
    0 & 0 & 0 & 0 & 1 & 0 & 0 & 0 \\
    0 & 0 & 1 & 0 & 1 & 0 & 0 & 0 \\
    0 & 1 & 0 & 0 & 1 & 0 & 0 & 0 \\
    0 & 1 & 1 & 0 & 1 & 0 & 0 & 0 \\
    1 & 0 & 0 & 0 & 1 & 0 & 0 & 0 \\
    1 & 0 & 1 & 1 & 0 & 1 & 0 & 0 \\
    1 & 1 & 0 & 1 & 0 & 1 & 0 & 0 \\
    1 & 1 & 1 & 1 & 0 & 1 & 0 & 0 \\
    \midrule
    \multicolumn{4}{l}{$c_i(a,b,c)$:} &
      $\overline{a\land (b \lor c)}$ &
      $a\land (b \lor c)$ &
      $0$ &
      $0$ \\
    \bottomrule
  \end{tabular}

  \caption{Direct circuit for $a\land (b \lor c)$ and the corresponding outputs
  $c_i(a,b,c)$.}
  \label{fig:a AND (b OR c)}
\end{figure*}

\begin{figure*}[t]
  \centering

  \includegraphics[width=0.5\linewidth]{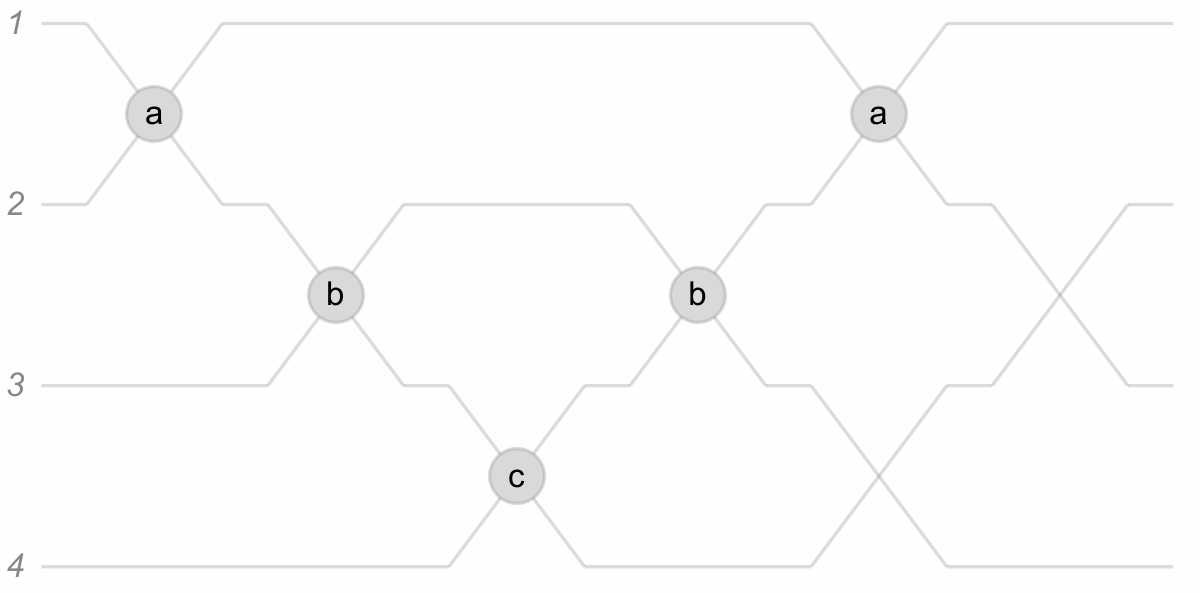}

  \vspace{1em}

  \begin{tabular}{cccc @{\quad} cccc}
    \toprule
    $a$ & $b$ & $c$ & $a\land (b \land c)$ &
      $c_1$ & $c_2$ & $c_3$ & $c_4$ \\
    \midrule
    0 & 0 & 0 & 0 & 1 & 0 & 0 & 0 \\
    0 & 0 & 1 & 0 & 1 & 0 & 0 & 0 \\
    0 & 1 & 0 & 0 & 1 & 0 & 0 & 0 \\
    0 & 1 & 1 & 0 & 1 & 0 & 0 & 0 \\
    1 & 0 & 0 & 0 & 1 & 0 & 0 & 0 \\
    1 & 0 & 1 & 0 & 1 & 0 & 0 & 0 \\
    1 & 1 & 0 & 0 & 1 & 0 & 0 & 0 \\
    1 & 1 & 1 & 1 & 0 & 1 & 0 & 0 \\
    \midrule
    \multicolumn{4}{l}{$c_i(a,b,c)$:} &
      $\overline{a\land (b \land c)}$ &
      $a\land (b \land c)$ & $0$ &
      $0$ \\
    \bottomrule
  \end{tabular}

  \caption{Direct circuit for $a\land (b \land c)$ and the corresponding outputs
  $c_i(a,b,c)$.}
  \label{fig:a AND (b AND c)}
\end{figure*}

\begin{figure*}[t]
  \centering

  \includegraphics[width=0.5\linewidth]{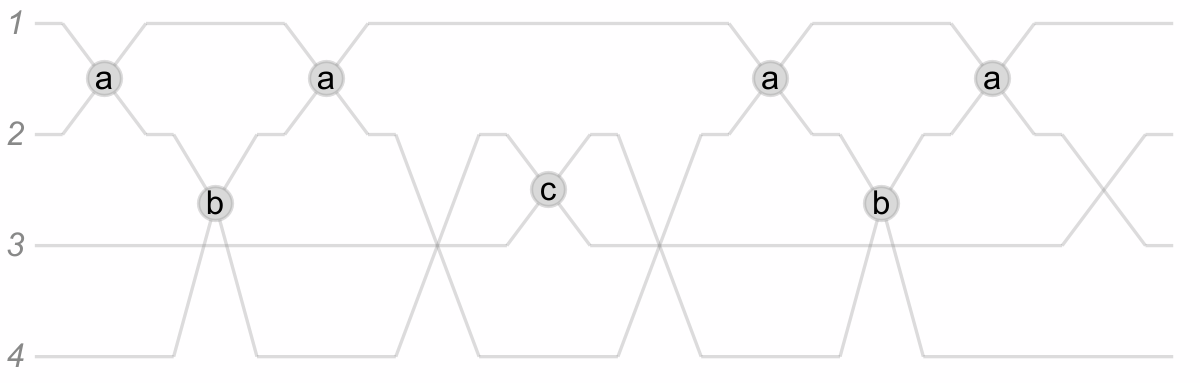}

  \vspace{1em}  

\begin{tabular}{cccc @{\quad} c @{\quad} cccc} 
\toprule
$a$ & $b$ & $c$ & $(a\land b) \land c$ & $c_1$ & $c_2$ & $c_3$ & $c_4$ & $c_5$ \\
\midrule
0 & 0 & 0 & 0 & 1 & 0 & 0 & 0 & 0 \\
0 & 0 & 1 & 0 & 1 & 0 & 0 & 0 & 0 \\
0 & 1 & 0 & 0 & 1 & 0 & 0 & 0 & 0 \\
0 & 1 & 1 & 0 & 1 & 0 & 0 & 0 & 0 \\
1 & 0 & 0 & 0 & 1 & 0 & 0 & 0 & 0 \\
1 & 0 & 1 & 0 & 1 & 0 & 0 & 0 & 0 \\
1 & 1 & 0 & 0 & 1 & 0 & 0 & 0 & 0 \\
1 & 1 & 1 & 1 & 0 & 1 & 0 & 0 & 0 \\
\midrule
\multicolumn{4}{l}{$c_i(a,b,c)$:} & $\overline{(a\land b) \land c}$ & $(a\land b) \land c$ & $0$ &  $0$ &  \
$0$ \\
\bottomrule
\end{tabular}

  \caption{Attempt to construct the direct circuit for $(a\land b)\land c$ and the corresponding outputs
$c_i(a,b,c)$. This configuration is the only one for which the truth
table does not correspond to the intended composition.}
  \label{fig:(a AND b) AND c}
\end{figure*}

\begin{figure*}[t]
  \centering

  \includegraphics[width=0.55\linewidth]{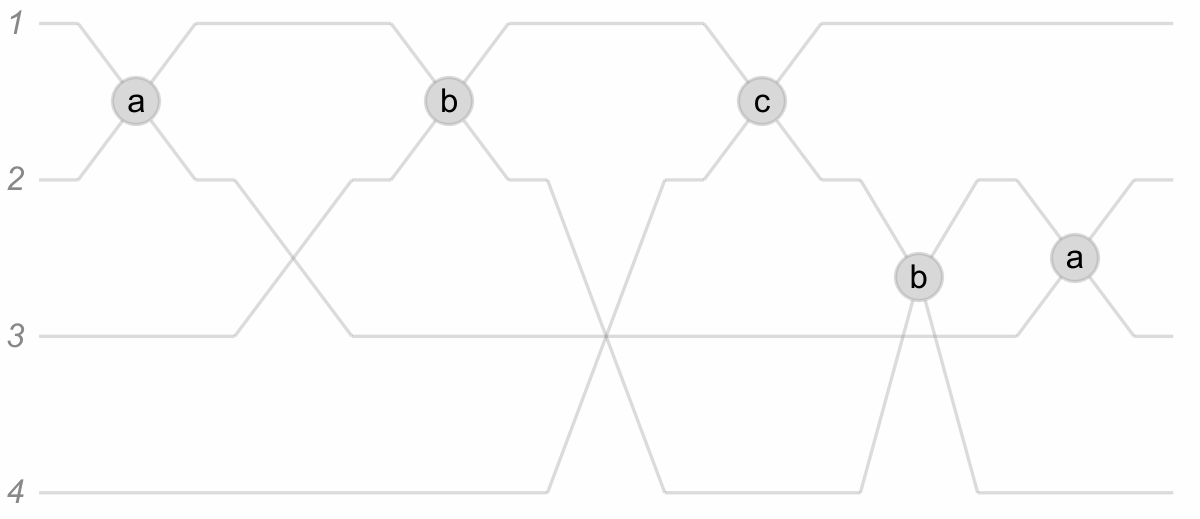}

  \vspace{1em}

  \begin{tabular}{cccc @{\quad} c @{\quad} ccc}
    \toprule
    $a$ & $b$ & $c$ & $a\lor (b \lor c)$ &
      $c_1$ & $c_2$ & $c_3$ & $c_4$ \\
    \midrule
    0 & 0 & 0 & 0 & 1 & 0 & 0 & 0 \\
0 & 0 & 1 & 1 & 0 & 1 & 0 & 0 \\
0 & 1 & 0 & 1 & 0 & 1 & 0 & 0 \\
0 & 1 & 1 & 1 & 0 & 1 & 0 & 0 \\
1 & 0 & 0 & 1 & 0 & 1 & 0 & 0 \\
1 & 0 & 1 & 1 & 0 & 1 & 0 & 0 \\
1 & 1 & 0 & 1 & 0 & 1 & 0 & 0 \\
1 & 1 & 1 & 1 & 0 & 1 & 0 & 0 \\
    \midrule
    \multicolumn{4}{l}{$c_i(a,b,c)$:} &
      $\overline{a\lor (b \lor c)}$ &
      $a\lor (b \lor c)$ & $0$ &
      $0$ \\
    \bottomrule
  \end{tabular}

  \caption{Direct circuit for $a\lor (b \lor c)$  and the corresponding outputs
  $c_i(a,b,c)$.}
  \label{fig:a OR (b OR c)}
\end{figure*}

\begin{figure*}[t]
  \centering

  \includegraphics[width=0.55\linewidth]{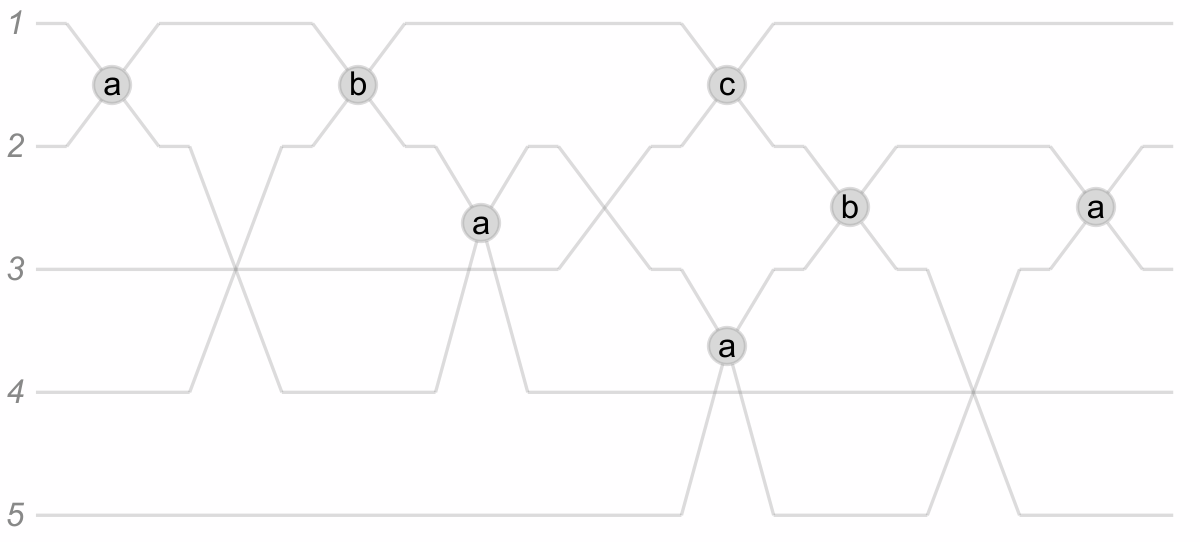}

  \vspace{1em}

\begin{tabular}{cccc @{\quad} c @{\quad} cccc} 
\toprule
$a$ & $b$ & $c$ & $(a\lor b) \lor c$ & $c_1$ & $c_2$ & $c_3$ & $c_4$ & $c_5$ \\
\midrule
0 & 0 & 0 & 0 & 1 & 0 & 0 & 0 & 0 \\
0 & 0 & 1 & 1 & 0 & 1 & 0 & 0 & 0 \\
0 & 1 & 0 & 1 & 0 & 1 & 0 & 0 & 0 \\
0 & 1 & 1 & 1 & 0 & 1 & 0 & 0 & 0 \\
1 & 0 & 0 & 1 & 0 & 1 & 0 & 0 & 0 \\
1 & 0 & 1 & 1 & 0 & 1 & 0 & 0 & 0 \\
1 & 1 & 0 & 1 & 0 & 1 & 0 & 0 & 0 \\
1 & 1 & 1 & 1 & 0 & 1 & 0 & 0 & 0 \\
\midrule
\multicolumn{4}{l}{$c_i(a,b,c)$:} & $\overline{(a\lor b) \lor c}$ & $(a\lor b) \lor c$ & $0$ &  $0$ &  \
$0$ \\
\bottomrule
\end{tabular}
  \caption{Direct circuit for $(a\lor b) \lor c$  and the corresponding outputs
  $c_i(a,b,c)$.}
  \label{fig:(a OR b) OR c}
\end{figure*}

\clearpage
\twocolumngrid

\section*{Funding}

This work was founded by Germany’s Excellence Strategy within the Cluster of Excellence PhoenixD 
(EXC 2122, Project ID 390833453).

 \section*{Data availability}
 \label{sec:avail-data-mater}

 All data supporting the findings of this study are available within
 the paper and its Supplementary Information.

 \section*{Code availability}
 \label{sec:avail-data-mater}

 The Mathematica package implementing the Directed Logic framework and all algorithms described in this work is publicly available at GitHub:
\url{https://github.com/ihar-babushkin/DirectedLogic}
The repository also contains the notebook used to reproduce the figures and truth tables presented in this paper.

\bibliography{optics}

\end{document}